# Hexagonal Rare-Earth Manganites as Promising Photovoltaics and Light Polarizers

*Xin Huang[1,2], Tula R. Paudel[1,*], Shuai Dong[2] and Evgeny Y. Tsymbal[1]*

[1]Department of Physics and Astronomy & Nebraska Center for Materials and Nanoscience

University of Nebraska, Lincoln, Nebraska 68588, USA

[2]Department of Physics, Southeast University, Nanjing 211189, China

Ferroelectric materials possess a spontaneous electric polarization and may be utilized in various technological applications ranging from non-volatile memories to solar cells and light polarizers. Recently, hexagonal rare-earth manganites, $h$-RMnO$_3$ (R is a rare-earth ion) have attracted considerable interest due to their intricate multiferroic properties and improper ferroelectricity characterized by a sizable remnant polarization and high Curie temperature. Here, we demonstrate that these compounds can serve as very efficient photovoltaic materials and, in addition, possess remarkable optical anisotropy properties. Using first-principles methods based on density-functional theory and considering $h$-TbMnO$_3$ as a representative manganite, we predict a strong light absorption of this material in the solar spectrum range, resulting in the maximum light-to-electricity energy conversion efficiency up to 33%. We also predict an extraordinary optical linear dichroism and linear birefringence properties of $h$-TbMnO$_3$ in a broad range of optical frequencies. These results uncover the unexplored potential of hexagonal rare-earth manganites to serve as photovoltaics in solar cells and as absorptive and birefringent light polarizers.

PACS number(s): 78.20.Ci, 78.20.Fm, 88.40.H-, 72.40.+w

# I. Introduction

Photovoltaic effects in ferroelectrics have been known for a long time.[1] The photovoltaic effect involves absorption of light, generating electrons and holes due to the photoelectric effect, and then separating them to generate a voltage. The photovoltaic effect in ferroelectrics relies on the internal electric field induced by the spontaneous electric polarization in the bulk absorber layer.[2] This internal electric field allows separating the photo-excited charge carriers, eliminating the need for an external bias to produce a potential drop in the interface-limited space charge region of a conventional semiconductor solar cell.[3] Ferroelectric photovoltaic materials also make possible to realize above-band-gap open-circuit voltage providing a large photovoltaic efficiency. The latter has been demonstrated using multiferroic perovskite $BiFeO_3$ thin films with structurally controlled steps of the electrostatic potential at nanometer-scale domain walls.[4,5,6]

Another kind of perovskite materials – organic halide perovskites, such as $CH_3NH_3PbI_3$, has recently fostered significant interest due to the improvement of efficiency of solar cells based on these materials up to 20% in last three years – an advance which was a matter of decades for other types of solar cells. [7,8] Organic halide perovskites have a long electron-hole diffusion length and low non-radiative recombination rates compared to other thin-film polycrystalline semiconductors.[9,10] In addition, they possess several desirable properties for a photovoltaic material such as strong optical absorption under light illumination, a direct band gap of appropriate magnitude, small effective masses both for electrons and holes and benign defect properties. [11,12,13,14] There are also indications that organic halide perovskites are ferroelectric, which may be used for efficient separation of elections and holes.[15] Unfortunately, organic perovskites have limited structural stability, which could hamper outdoor applications.

Inorganic ferroelectric materials (mostly oxide perovskites), on the other hand, often exhibit an electronic band gap being too large for an efficient absorber of the solar radiation. For example, the band gap of $BiFeO_3$ is about 2.7 eV which prevents the absorption in a large part of the visible spectrum. Some other conventional ferroelectric perovskite oxides have band gaps even higher than that of BiFeO$_3$. For example, band gaps of BaTiO$_3$, PbTiO$_3$, and KNbO$_3$ are 3.2 eV, 3.4 eV, and 4 eV, respectively. In fact, earlier efforts to make use of these materials as photovoltaics yielded a very small efficiency due to low optical absorption.[16] More recently, aliovalent doping in KNbO$_3$ [17,18] and modification of the site population or cation ratio in multiferroic Bi$_2$FeCrO$_6$ [19] is used to engineer absorption properties of ferroelectric materials and to enhance their light-to-electricity conversion efficiency. In particular, the power conversion efficiency in multilayer multiferroic Bi$_2$FeCrO$_6$ solar cells is found to reach 8.1% [19], which is attributed to the low band gap and intrinsic ferroelectricity of the compound.

In this paper, we explore optical properties and a photovoltaic effect in intrinsically narrow band gap oxide ferroelectric materials – hexagonal rare-earth manganites, *h*-RMnO$_3$ (R is a rare-earth ion). These materials have recently attracted a lot of excitement due to their improper structural ferroelectricity and its coupling to geometrically frustrated spins. The ferroelectric non-centrosymmetric *P6$_3$cm* phase of *h*-TbMnO$_3$ is energetically more favorable than the paraelectric centrosymmetric *P6$_3$/mmc* phase,[32] in which the polarization originates from the rotation and tilt of the MnO$_5$ polyhedra and anti-phase displacement of Tb and O atoms [20] similar to that found in other *h*-*R*MnO$_3$ compounds.[21,22]. This mechanism of ferroelectricity is different from that of orthorhombic TbMnO$_3$ where ferroelectricity is induced by spin frustration.[23]. At the low temperature, the antiferromagnetic order is mediated by a non-collinear in-plane Mn-O-Mn and collinear Mn-O-R-O-Mn super-exchange interactions induced by ferroelectric distortions of the rare-earth ions making these compounds multiferroic.[24] Furthermore, *h*-RMnO$_3$ materials undergo an isostructural transition with exceptionally large atomic displacements, resulting in an unusually strong magneto-elastic coupling.[25] The broken space inversion and time reversal symmetries allow a linear magnetoelectric coupling and produce coupled electric and magnetic domains.[26,27] These domains form a unique 'cloverleaf' topology pattern with six interlocked structural antiphase and ferroelectric domains merging into a vortex core.[28]

The optical properties of the rare-earth manganites appear to be equally exciting from the point of view of light-to-energy conversion as well as production of a polarized light. Their band gaps are of about 1.3 – 1.6 eV, which according to the well-known Shockley-Queisser criterion, are ideal for utilizing a large part of the visible-light solar radiation for optical excitation.[29] Additionally, these materials have a relatively large remnant ferroelectric polarization with a high Curie temperature that could provide a large intrinsic electric field needed for separating electron-hole pairs. The optical spectroscopy measurements show that these compounds have a sharp absorption edge and have highly anisotropic optical properties. [30,31,32,33]

In rare-earth series, manganites of larger rare-earth ions (La to Dy) crystallize in orthorhombic phase, whereas those of smaller rare-earth ions (Ho to Lu or Y) crystallize in hexagonal phase. Tb and Dy manganites can be epitaxially stabilized in hexagonal phase. Here, we consider hexagonal-TbMnO$_3$ (*h*-TbMnO$_3$) as a model system. Similar to other manganites, this material is a charge transfer insulator with a direct band gap of about 1.4 eV ideal for photovoltaic applications. In contrast to the bulk orthorhombic phase, *h*-TbMnO$_3$ shows a relatively large remnant ferroelectric polarization (8 μC/cm$^2$) and a high Curie temperature (> 590K).[34,20] This sizable remnant polarization is expected to provide a large intrinsic electric field needed for separating electron-hole pairs. High quality *h*-TbMnO$_3$ films can be grown on suitable substrates such as Pt (111)/Al$_2$O$_3$ (0001) and YSZ (111). [34]

We use the first-principles approach (see *First principles* for details) to study the photovoltaic properties of *h*-TbMnO$_3$ and compare these properties to those of CdTe, a commercial solar cell absorber, and BaTiO$_3$, a conventional large band gap ferroelectric. We find that *h*-TbMnO$_3$ has a high absorption coefficient with a particularly sharp onset, due to interband transitions between hybridized Mn *3d* ground and excited states. This leads to a very high light-to-electricity energy conversion efficiency with a maximum energy conversion efficiency reaching 33%. Furthermore, we demonstrate a remarkable anisotropy in the optical properties of this material, resulting in large linear dichroism and linear birefringence, which allow using *h*-TbMnO$_3$ as a light polarizer in the visible spectrum range.

## II. Results and Discussions

### A. The Optical properties and lattice structure

The atomic structure of the hexagonal phase of TbMnO$_3$ (space group $P6_3cm$) is characterized by Mn$^{3+}$ ions surrounded by five O$^{2-}$ ions creating a trigonal bipyramidal environment (Fig. 1a and 1b). The layers of MnO$_5$ bipyramids are interleaved by the planes of Tb$^{3+}$ ions. All Mn$^{3+}$ ion are symmetrically equivalent and occupy 6$c$ Wyckoff position, while the symmetry of Tb$^{3+}$ ions is lower with two of them in 2$a$ and the other one in 4$b$ Wyckoff position. Oxygen are bonded with Tb and Mn and occupy 2$a$, 6$c$ and 4$b$ Wyckoff's positions. Under the $P6_3cm$ trigonal symmetry the Mn 3$d$ orbitals are split into two doublets, $e'(x^2-y^2, xy)$ and $e''(yz, zx)$, and one singlet $a'_1(z^2)$.[35] The orbitals are filled according to the strong Hund's rule coupling. For the Mn$^{3+}$(3$d^4$) configuration, both the $xz$ and $yz$ orbitals, which are directed between the oxygens, and the in-plane $xy$ and $x^2-y^2$ orbitals are singly occupied, whereas the $z^2$ orbital pointing to the two apical oxygens is empty (Fig. 1c). Therefore, the two doublet $e''$ and $e'$ states form the valence band while the singlet $a'_1$ forms the conduction band.

Fig. 2 shows the calculated optical absorption spectra of $h$-TbMnO$_3$ in comparison with those of CdTe, the most common solar cell absorber, and BaTiO$_3$, a prototypical ferroelectric with a large band gap. It is seen from the figure that the optical absorption of $h$-TbMnO$_3$ exhibits strong anisotropy with respect to the relative orientation of the polarization of light and the crystal lattice. This polarization anisotropy results from anisotropy of the $h$-TbMnO$_3$ dielectric tensor $\varepsilon(\omega)$ (see Fig. 3). For hexagonal crystals, such as $h$-TbMnO$_3$, the dielectric tensor is non diagonal, i.e. $\varepsilon_{xx} = \varepsilon_{yy} \neq \varepsilon_{zz}$. We denote the two different components of the dielectric tensor using the orientation of electric field relative to the optical axis.[36] The dielectric component for e $\perp$ c is $\varepsilon_{\perp} = \varepsilon_{xx} = \varepsilon_{yy}$ and that for e ∥ c is $\varepsilon_{\parallel} = \varepsilon_{zz}$, so we label the optical absorption coefficient for the light polarization in the *ab*-plane as $\alpha_\perp$ and along the *c*-axis as $\alpha_\parallel$. Especially strong difference between $\alpha_\perp$ and $\alpha_\parallel$ is seen in the energy window 1.5 eV < *E* < 2 eV where $\alpha_\perp > \alpha_\parallel$ (Fig. 2). In contrast to the hexagonal TbMnO$_3$, CdTe is cubic and thus the three components of its dielectric tensor are equal, i.e. $\varepsilon_{xx} = \varepsilon_{yy} = \varepsilon_{zz}$. BaTiO$_3$ is tetragonal at room temperature, but the tetragonal distortion is small, so that the optical absorption spectrum is also nearly isotropic.

The pronounced asymmetry in the optical absorption of *h*-TbMnO$_3$ between polarizations of light along the *c*-axis and in the *ab*-plane leads to a strong linear dichroism for this material in the visible spectral region. Fig. 4 shows the calculated linear dichroism $\Delta k = k_\perp - k_\parallel$ (red curve), along with the linear birefringence $\Delta n = n_\perp - n_\parallel$ (blue curve). Large positive values of $\Delta k$ for photon energy less than 2.8 eV (except a narrow region around 2.5 eV) reflect a strong anisotropy of the absorption coefficient ($\alpha_\perp > \alpha_\parallel$) in this energy range. The linear dichroism changes sign at about *E* = 3 eV. A sizable birefringence is indicative of a refractive index that depends on the polarization and propagation direction of light. Both, linear dichroism and linear birefringence can be exploited for making light polarizers and optical filters. Thus, *h*-TbMnO$_3$ single crystals can be used both for absorptive and birefringent polarizers of light in the visible spectrum range.

A particularly important characteristic for efficient photovoltaic materials is a sharp absorption edge and a large magnitude of optical absorption. As seen from Fig. 2, this is the case for *h*-TbMnO$_3$ whose optical absorption ($\alpha_\perp$) exhibits a remarkably sharp onset at photon energy of 1.4 eV equal to the experimental band bap of *h*-TbMnO$_3$ and maintains a large magnitude (~10$^5$ cm$^{-1}$) throughout the visible frequency range. We find that $\alpha_\perp$ for *h*-TbMnO$_3$ is about two times larger than $\alpha$ for CdTe at photon energy *E* < 3.0 eV (except a narrow energy window near 2.5 eV), and much larger than $\alpha$ for BaTiO$_3$ being transparent in the whole visible spectrum range (*E* < 3.2 eV). The other notable features of the $\alpha_\perp$ spectrum for $h$-TbMnO$_3$ include a plateau of large magnitude ($10^5$ cm$^{-1}$) in range of photon energies 1.6 eV < $E$ < 2.3 eV, an absorption dip at $E$ = 2.5 eV and a prominent peak at $E$ = 2.7 eV. Even though the absorption coefficient is somewhat reduced at $E \sim 2.5$ eV and $E > 2.7$ eV, the absolute value still remains sizable (> $10^4$ cm$^{-1}$).

**B. Electronic Structure of $h$-TbMnO$_3$**

These features of the absorption spectra can be understood by analyzing the electronic band structure of $h$-TbMnO$_3$. Fig. 5 shows the calculated band structure along high-symmetry directions of the $h$-TbMnO$_3$ Brillouin zone at energies around the band gap. Both the valence and conduction bands are composed of the Mn $3d$ and O $2p$ states with varying degree of their hybridization. Tb states lie at energies far away from the relevant energy range. The lowest laying conduction bands consist of the three Mn $a'_1$ states originating from the same spin orbitals split due to the in-plane and out-of-plane hybridizations (Fig. 6). The highest valence bands are composed of the hybridized Mn $e'$, Mn $e''$ and O $2p$ states. These relatively non-dispersive bands provide a high density of states that may be responsible for a strong optical absorption if the optical transitions are allowed by the dipole selection rules. According to the band structure alone, we expect very strong optical absorption along the A–Γ, K–H, and L–M directions parallel to the $c$ axis, which are labeled as Δ, $P$, and $U$, respectively (Fig. 4 (a) inset).

To determine the allowed optical transitions, we analyze the electric dipole matrix elements along the three high-symmetry lines within the first Brillouin zone. The results are given in Table 1. The selections rules for the optical transition matrix elements in conjunction with the band structure in Fig. 4 explain the features in the $\alpha_\perp$ and $\alpha_\parallel$ absorption spectra shown in Fig. 2.

At lower photon energies, the optical absorption for polarization perpendicular to the *c* direction ($\alpha_\perp$) is controlled by the $\langle a'_1 | q_x | e' \rangle$ and $\langle a'_1 | q_y | e' \rangle$ dipole matrix elements. As seen from Table 1, these transitions are allowed along the Γ–A direction, where we see rather flat $e'$ bands at the top of the valence band and $a'_1$ bands at the bottom of the conduction band (Fig. 5). These bands are separated by the energy gap of about 1.4 eV, have high density, and are responsible for the onset of the optical absorption $\alpha_\perp$ at $E$ = 1.4 eV (Fig. 2). With increasing the photon energy the $\alpha_\perp$ absorption remains large, due to the $\langle a'_1 | q_x | e' \rangle$ and $\langle a'_1 | q_y | e' \rangle$ transitions being allowed across the whole Brillouin zone and available valence and conduction bands over the whole range of photon energies, except the dip at around $E$ = 2.5 eV caused by the splitting of the $a'_1$ excited states (Fig. 5). The peak at $E$ = 2.7 eV is associated with the optical transitions from the upper $e'$ valence bands to the top less dispersive $a'_1$ conduction band. Contrary to the $\alpha_\perp$ absorption, there is no optical absorption for polarization parallel to the *c* direction ($\alpha_\parallel$) in the range of photon energies 1.4 eV < $E$ < 2 eV. This is due to the fact that the dipole transition $\langle a'_1 | q_z | e' \rangle$ responsible for $\alpha_\parallel$ absorption is forbidden along the Γ–A line (Table 1). The onset of the $\alpha_\parallel$ absorption occurs at the higher energies, when moving away from the A point along the A–L line we approach the L point and then the L–M line (Fig. 2) where the $\langle a'_1 | q_z | e' \rangle$ dipole transition is allowed (Table 1). The respective conduction bands lie at 2.2 eV and higher energies and provide a continuing increase in the $\alpha_\parallel$ absorption with the increasing photon energy.

## C. Photovoltaic Energy Conversion Efficiency of *h*-TbMnO$_3$

We calculate the maximum photovoltaic energy conversion efficiency, i. e. the ratio of electrical energy output from the solar cell to input energy from the sun, which is the most commonly used parameter to compare the performance of one solar cell to another. The conventional approach to evaluate the photovoltaic conversion efficiency is the well-known Shockley-Queisser (SQ) method.[29] Within this method, the optical absorption spectrum is assumed to be a step function with onset at the band gap energy. As the result, the efficiency is just a function of the band gap and hence does not depend on details of the absorption spectrum. The *SQ* efficiency vs band gap energy curve peaks at around 1.4 eV, incidentally the band gap of *h*-TbMnO$_3$, thereby suggesting that this material could be an efficient solar-cell absorber. However, the photovoltaic efficiency depends upon other materials specific properties, such as details of the absorption spectrum, carrier recombination, and thickness of the absorber layer, which are not included in the SQ approach. Therefore, we use a spectroscopic limited maximum efficiency method to evaluate the photovoltaic energy conversion efficiency.[37] This method employs a thermodynamic approach taking into account the specific shape of the optical absorption spectrum and the energy dependence of radiative recombination losses (see *Methods* for details).

Fig. 7 shows the calculated photovoltaic cell efficiencies for *h*-TbMnO$_3$, CdTe and BaTiO$_3$ as a function of film thickness *L*. We see that the efficiency increases with increasing *L*. At small *L*, the efficiency is small due to small absorption. At large *L*, the photon absorption reaches maximum independent of *L* and is different for different materials. In particular, for *h*-TbMnO$_3$, we find the maximum conversion efficiency to be as large as 33%. For a typical absorber layer thickness of *L* = 0.5 μm, the calculated efficiencies for *h*-TbMnO$_3$ is 32% compared to 27% for CdTe and 2% for BaTiO$_3$. We also obtain other thickness-dependent solar cell parameters, such as the short-circuit current density $J_{sc}$, open-circuit voltage $V_{oc}$ and fill factor (Fig. 8). The experimentally measured $J_{sc}$ = 26 mA/cm$^2$ and $V_{oc}$ = 0.87 eV in CdTe are close to our calculated theoretical limits for this material: $J_{sc}$ = 23 mA/cm$^2$ and $V_{oc}$ = 1.3 eV at 0.5 m, indicating that our method is reliable.[38] Compared to CdTe, the calculated limits for *h*-TbMnO$_3$ are $J_{sc}$ = 30.9 mA/cm$^2$ and $V_{oc}$ = 1.16 eV at 0.5 μm, showing that *h*-TbMnO$_3$ has the capability to realize a high short-circuit current. We point out that the photovoltaic performance of *h*-TbMnO$_3$ is comparable with that of the CH$_3$NH$_3$PbX$_3$-based solar cell, as the experimentally achieved values of $J_{sc}$ and $V_{oc}$ are 22.4 mA/cm$^2$ and 0.92 eV, respectively, for film thickness of 0.35 μm.[39] Hence, due to a proper band gap, a very large absorption coefficient in the visible spectrum range, and a relatively large remnant ferroelectric polarization which might be exploited to separate electron-hole pairs, we predict that $h$-TbMnO$_3$ is a very promising material for high efficiency light-to-electricity conversion in solar cell devices.

## III. Conclusion

We envision similar photovoltaic and light-polarization properties in other rare-earth manganites. Manganites with smaller rare earth ions, such as Ho, Er, Tm, Yb, Lu or Y, are stable in the hexagonal bulk phase. These compounds have a band gap ranging from about 1.3 to 1.6 eV, ideal for effective absorption of the visible-light solar radiation and photo-excitation. The electronic structure of the valence and conduction bands and the optical transition selection rules are determined by symmetry and therefore are similar to those of $h$-TbMnO$_3$. This suggests that the hexagonal rare-earth manganites have a large anisotropic optical absorption with a sharp onset. All these materials are ferroelectric with a large spontaneous polarization and the ferroelectric transition temperature well above room temperature. The presence of internal electric field driven by ferroelectricity allows an efficient separation of the photo-excited electron-hole pairs, eliminating the need for an external bias in a solar cell. Additionally alloying of the rare earth ions in these compounds can be used to tune the required material properties including the optical absorption and mobility by modifying electronic structure of band edges. The mobility of holes in these compounds are expected to be low due to non-dispersive valance band mainly made of Mn-d orbitals. Further, the carrier conductivity could suffer loss by carrier localization coupled to local deformations. All these materials however, are stable at high temperatures and are iso-structural to common polar transparent conducting oxides, such as ZnO, which not only helps to retain the polarization and the intrinsic electric field, but also offers high quality growth.

We hope therefore that our theoretical predictions will stimulate experimental studies to uncover the unexplored potential of hexagonal rare-earth manganites as photovoltaics in solar cells and as absorptive and birefringent light polarizers.

**Acknowledgements**

This work was supported by the National Science Foundation (NSF) under the Nebraska MRSEC program (Grant No. DMR-1420645). The research at Southeast University was funded by the National Science Foundation of China (Grant No. 51322206). X. H. acknowledges the Fundamental Research Funds for the Central Universities and Jiangsu Innovation Projects for Graduate Student (Grant No. 51471085). Computations were performed utilizing the Holland Computing Center at the University of Nebraska. The authors acknowledge discussion with Prof. Sitaram Jaswal regarding the group theory analysis of optical selection rules.

**Supporting Information**

Supporting Information is available online PRB or form authors.

# Appendix

## I. Methods

### A. First-principles calculations

We use spin polarized density-functional theory (DFT) calculations based on the generalized gradient approximation (GGA) with Perdew-Burke-Ernzerhof (PBE) potentials, as implemented in Vienna *ab initio* Simulation Package (VASP).[40,41] We consider $h$-TbMnO$_3$ of $P6_3cm$ space group symmetry with experimental lattice constants $a = b = 6.27$ Å, $c = 11.46$ Å and relax the internal co-ordinates with the force convergence limit of 5 meV/atom. The plane-wave cutoff energy of 525 eV and a 7×7×4 Monkhorst-Pack $k$-point mesh centered at the Γ point for total-energy calculation are employed. We treat correlation effects beyond GGA at a semi-empirical GGA+$U$ level within a rotationally invariant formalism for the Mn 3$d$-orbitals. In order to get a reasonable experimental lattice parameters, polarization and overall electronic structure, we apply $U$ = 1.5 eV on the Mn 3$d$-orbitals.[19] While the calculated band gap of $h$-TbMnO$_3$ is 0.84 eV, which is smaller than the experimental band gap of 1.4eV due to the well-known underestimation of electron-electron correlations within the local density approximation. A collinear antiferromagnetic structure is assumed in the calculations, where the magnetic moments of two Mn atoms in the plane are pointing opposite to the magnetic moment of the third Mn atom in that plane (Fig. 6(a) and 6(c)). Mn moments within the two Mn planes in the unit cell are opposite making the net magnetic moment zero (Fig. 6(b)). This frustrated magnetic order has lower energy compared to a pure A-type antiferromagnetic order.[42]

**B. Optical properties**

The absorption spectra are calculated using the energy dependent dielectric functions. Within the one-electron picture, the imaginary part of the dielectric function $\varepsilon_2(\omega)$ is obtained from the following equation:[43]

$$\varepsilon_2(\omega) = \frac{4\pi^2 e^2}{\Omega} \lim_{q \to 0} \frac{1}{q^2} \sum_{c,\upsilon,\mathbf{k}} 2w_{\mathbf{k}} \delta(E_c - E_\upsilon - \omega) |\langle c | \mathbf{e} \cdot \mathbf{q} | \upsilon \rangle|^2. \qquad (1)$$

According to Eq. 1, $\varepsilon_2(\omega)$ is determined by the integrated optical transitions $\langle c | \mathbf{e} \cdot \mathbf{q} | \upsilon \rangle$ from the valence states ($\upsilon$) to the conduction states ($c$), where **e** is the polarization direction of the photon and **q** is the electron momentum operator. The integration over the **k** is performed by summation over special k-points with corresponding weighting factor $w_{\mathbf{k}}$. The real part of the dielectric function $\varepsilon_1(\omega)$ is obtained from the imaginary part $\varepsilon_2(\omega)$ using the Kramers-Krönig transformation, which are shown in Fig. 3. $\varepsilon_1(\omega)$ and $\varepsilon_2(\omega)$ are then used to calculate the optical absorption coefficient $\alpha(\omega)$, the refractive index $n(\omega)$ and the extinction coefficient $k(\omega)$, using the following relations:[44]

$$\alpha(\omega) = \sqrt{2}\frac{\omega}{c}[\sqrt{\varepsilon_1^2(\omega)+\varepsilon_2^2(\omega)}-\varepsilon_1(\omega)]^{1/2}, \qquad (2)$$

$$n(\omega) = \frac{1}{\sqrt{2}}[\sqrt{\varepsilon_1^2(\omega)+\varepsilon_2^2(\omega)}+\varepsilon_1(\omega)]^{1/2}, \qquad (3)$$

$$k(\omega) = \frac{1}{\sqrt{2}}[\sqrt{\varepsilon_1^2(\omega)+\varepsilon_2^2(\omega)}-\varepsilon_1(\omega)]^{1/2}. \qquad (4)$$

In the calculations of optical properties, we use the scissor operator, $\Delta$, which rigidly shifts the eigenvalues of the conduction bands upward to correct the calculated band gap.[45] A similar approach is used for CdTe (0.79 eV calculated and 1.5 eV experimental band gaps) and BaTiO$_3$ (2.45 eV calculated and 3.2 eV experimental band gaps), which we employ as a benchmark for optical and photovoltaic properties.

## C. Photovoltaic properties

A photovoltaic cell can be approximated by equivalent ideal diode illuminated under the incident photon flux $I_{sun}$. When a load resistor is connected to an illuminated photovoltaic cell, the total current can be calculated as follows:[46]

$$J = J_{sc} - J_D = J_{sc} - J_0(e^{qV/kT}-1). \qquad (5)$$

Here, the first term $J_{sc}$ is the short-circuit current density driven by optical generation, which is given by $J_{sc} = e\int_0^\infty a(E)I_{sun}(E)dE$, where $I_{sun}(E)$ is the solar radiation flux and $a(E)$ is the photon absorptivity (Fig. 8(a)). The absorptivity is given by $a(E) = 1-e^{-2\alpha(E)L}$ and depends on the absorption coefficient $a(E)$ and absorber thickness $L$. For $h$-TbMnO$_3$, we assume that the absorption coefficient is given by an average over light polarizations so that $\alpha = (2\alpha_\perp + \alpha_\parallel)/3$. The second term in Eq. 5, $J_D = J_0(e^{qV/kT}-1)$, is the dark current density that depend on the electron-hole recombination current density, $J_0 = J_0^{nr} + J_0^{r}$, involving both the non-radiative $J_0^{nr}$ and radiative $J_0^r$ processes, at temperature $T$ and voltage $V$. The non-radiative recombination is dominant for indirect band gap compounds and typically is negligible for direct band gap compounds. The latter is the case for $h$-TbMnO$_3$ where the radiative recombination is the dominating loss mechanism. We calculate the radiative recombination rate using a detailed-balance principle assuming the rates of emission and absorption through cell surfaces to be equal.[47] A current due to the black body radiation absorption in the dark can be used to estimate the emission current. This current is obtained similar to $J_{sc}$, but using the black body radiation flux $I_{bb}$, so that $J_0^r = e\int_0^\infty a(E)I_{bb}(E,T)dE$. Thus, by knowing the absorption coefficient we can calculate $J_{sc}$ and $J_0$. The open circuit voltage ($V_{oc}$) shown in Fig. 8(b) is the voltage when $J_{sc}$ is zero. Finally, the maximum energy conversion efficiency is obtained from $\eta = P_{out}^{max}/P_{in}$. Here $P_{in}$ is the total incident solar power density and $P_{out}^{max}$ is the maximum electrical output power density, which can be evaluated by numerically maximizing the formula $P_{out} = J \cdot V = [J_{sc} - J_0(e^{qV/kT}-1)]V$. The fill factor (*FF*) is calculated from $FF = P_{out}^{max}/J_{sc}V_{oc}$ and shown in Fig. 8(c).

**Figures**

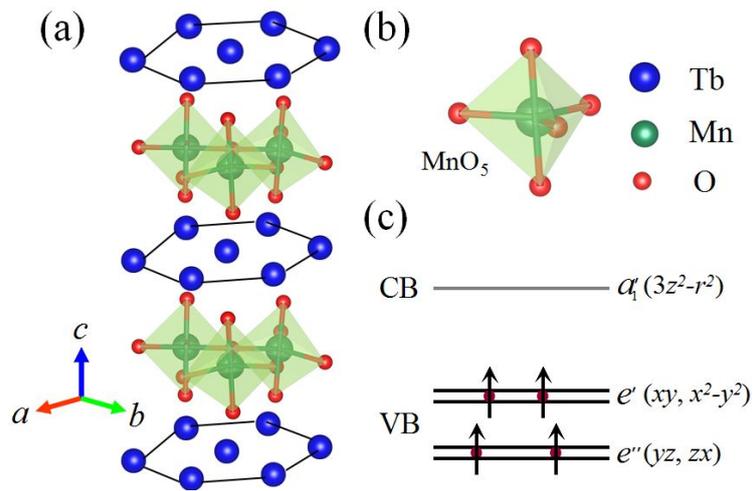

Fig. 1. (Color online) Crystal structure and $d$ orbitals splitting for $h$-TbMnO$_3$. (a) Crystal structure of bulk $h$-TbMnO$_3$ in space group $P6_3cm$. The blue, green and red spheres represent Tb, Mn and O atoms, respectively. **(b)** Mn-centered trigonal bipyramid formed of Mn$^{3+}$ and five O$^{2-}$ ions. **(c)** Schematic representation of the $d$ orbitals crystal field splitting under the MnO$_5$ triangular bipyramid environment. The two doublet states, $e''$ and $e'$, are occupied and form the valence band (VB), whereas the singlet $a'_1$ is unoccupied and forms the conduction band (CB).

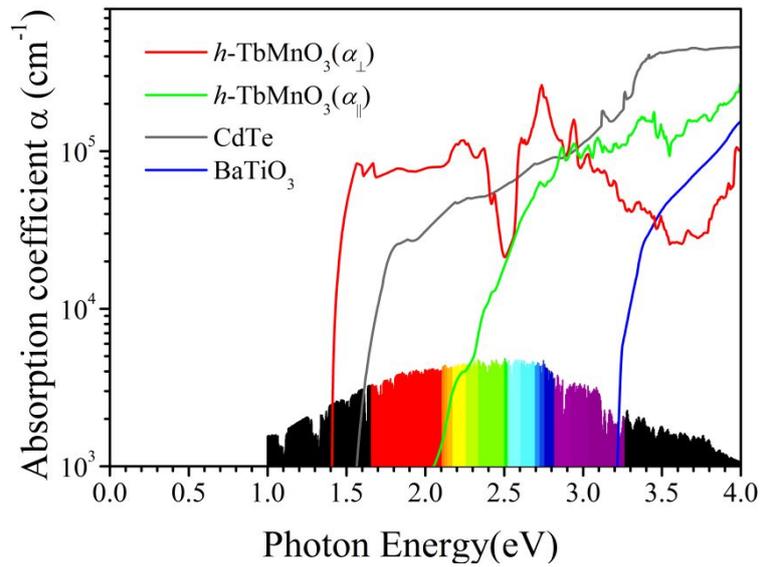

Fig. 2. (Color online) Optical absorption spectra. Solid lines of different color represent the calculated results for *h*-TbMnO$_3$ (red line for polarization of light in the *ab*-plane and green line for polarization of light out-of-plane along the *c*-axis), CdTe (black line) and BaTiO$_3$ (blue line). The standard AM1.5G solar spectrum is shown on bottom as a reference with different colors from red to purple indicating the visible light spectrum. [48]

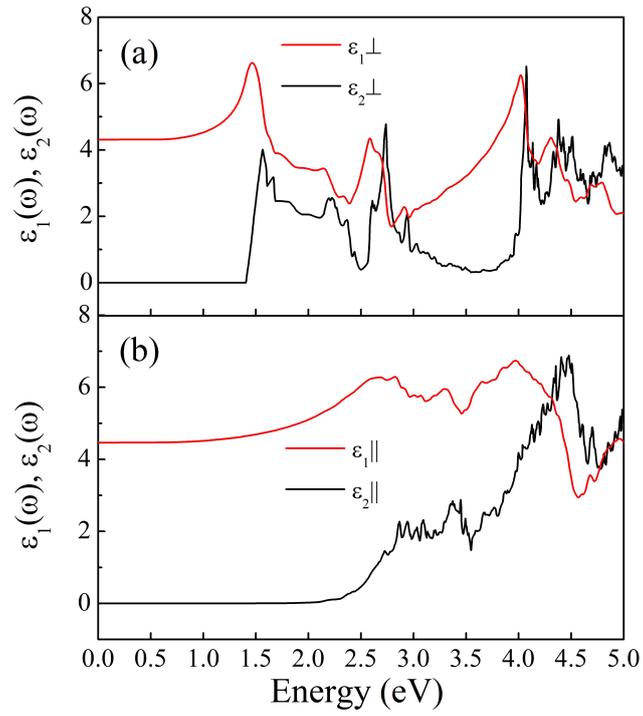

**Fig. 3.** (Color online) The calculated real (red curve) and imaginary (black curve) dielectric spectra $\varepsilon_\perp$ (a) and $\varepsilon_\parallel$ (b).

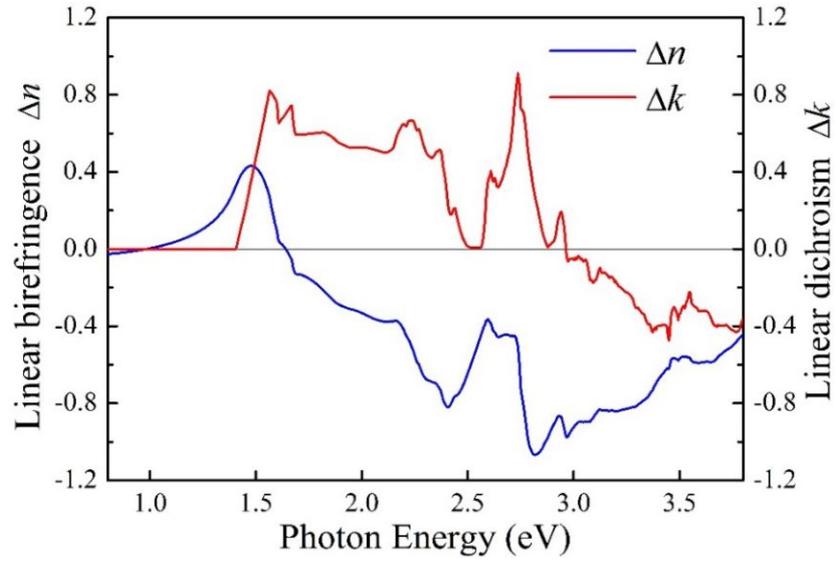

Fig. 4. (Color online) Anisotropic optical properties of $h$-TbMnO$_3$. Linear dichroism $\Delta k = k_\perp - k$ (red line) and linear birefringence $\Delta n = n_\perp - n$ (blue line).

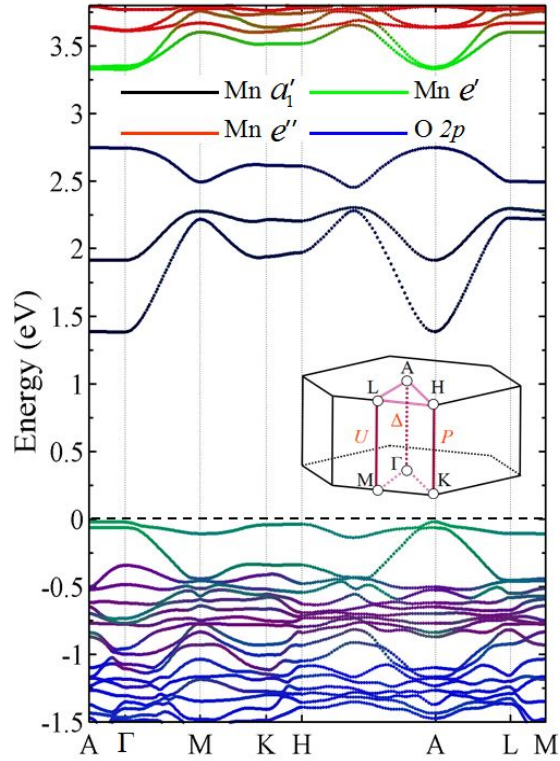

Fig. 5. (Color online) Electronic band structure of *h*-TbMnO$_3$. The bands are shown along lines connecting high-symmetry points in the Brillouin zone, A (0,0,1/2), Γ (0,0,0), M (1/2,0,0), K (2/3,1/3,0), H (2/3,1/3,1/2), L (1/2,0,1/2). The inset shows the irreducible part of the first Brillouin zone. Bands are colored according to their orbital character: Mn $a'_1$ (black), Mn $e'$ (green), Mn $e''$ (red), and O 2*p* (blue). The lower conduction bands consist of Mn $a'_1$ orbitals, whereas Mn $e'$, Mn $e''$ and O 2*p* orbitals contribute to the valence bands. The upper valence bands are dark green due to Mn $e'$ orbitals slightly intermixed with O 2*p* orbitals. The valence bands around -0.75 eV is purple due to strong hybridization between Mn $e''$ (red) and O 2*p* (blue) states. Bands at energies below -1 eV are blue showing that they largely consist of the O 2*p* states.

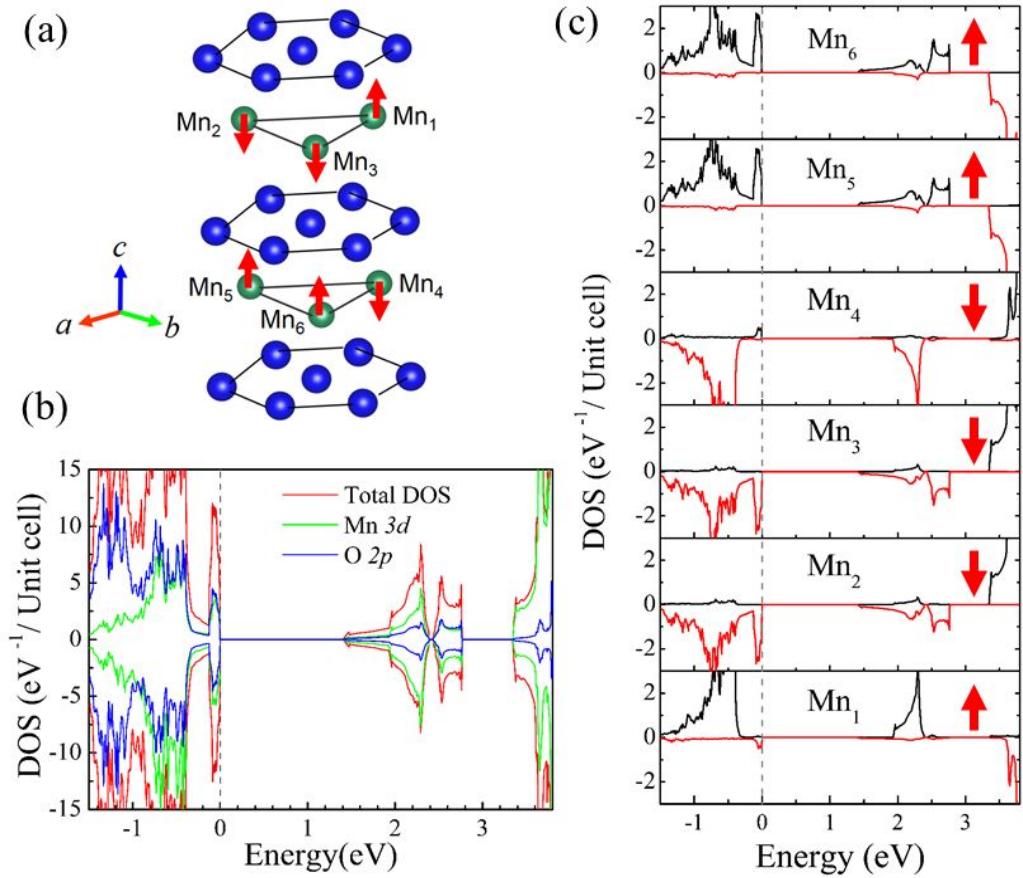

**Fig. 6.** (Color online) (a) Spin configuration of Mn atoms in TbMnO$_3$. (b) Total density of states (red curve) and projected density of states for Mn *3d* orbitals (green curve) and O *2p* orbitals (blue curve). (c) Mn site resolved density of states. The red arrows indicate the direction of Mn magnetic moments. The magnetic moments within the plane form a frustrated triangular antiferromagnetic structure. The net magnetic moment of the two adjacent Mn planes is zero.

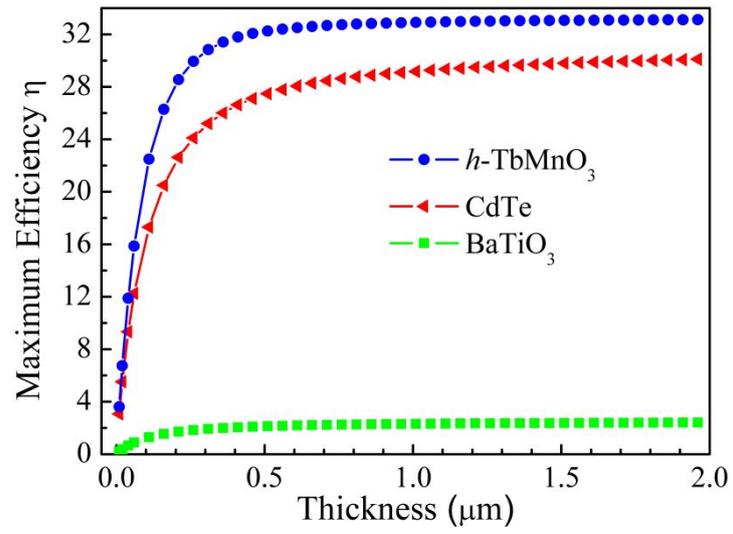

Fig. 7. (Color online) Calculated maximum photovoltaic energy conversion efficiency for *h*-TbMnO$_3$, CdTe, and BaTiO$_3$ as a function of absorber layer thickness.

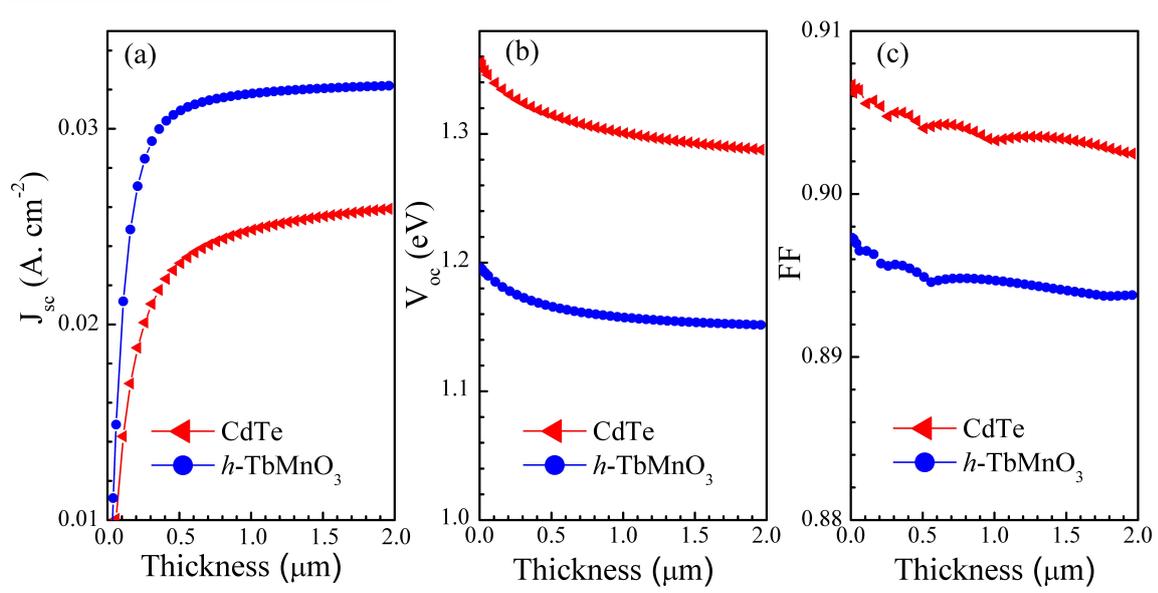

**Fig. 8.** (Color online) The calculated device parameters $J_{sc}$ (a), $V_{oc}$ (b), and fill factor (c) for *h*-TbMnO$_3$ (blue) ideal solar cell and CdTe (red) ideal solar cell as a function of absorber layer thickness.

**Table 1.** The electric dipole transitions between valence bands ($e''$ or $e'$) and conduction bands ($a'_1$) along high-symmetry directions in the irreducible part of the Brillouin zone of *h*-TbMnO$_3$ (inset in Fig. 5): Δ (Γ to A, 6*mm* symmetry), P ( K to H, 3*m* symmetry), and *U* (L to M, 2*mm* symmetry). The three components of the dipole moment **q** are listed separately. Allowed (forbidden) dipole transitions are indicated by A (F).

|  | Γ $\xrightarrow{\Delta}$ A (6*mm*) | K $\xrightarrow{P}$ H (3*m*) | L $\xrightarrow{U}$ M (2*mm*) |
|---|---|---|---|
| $\langle a'_1 \vert q_x \vert e' \rangle$ | A | A | A |
| $\langle a'_1 \vert q_y \vert e' \rangle$ | A | A | A |
| $\langle a'_1 \vert q_z \vert e' \rangle$ | F | F | A |
| $\langle a'_1 \vert q_x \vert e'' \rangle$ | F | A | F |
| $\langle a'_1 \vert q_y \vert e'' \rangle$ | F | A | F |
| $\langle a'_1 \vert q_z \vert e'' \rangle$ | F | F | F |